%% file: main.tex
\documentclass[conference]{IEEEtran}
\IEEEoverridecommandlockouts

\usepackage{cite}
\usepackage{amsmath,amssymb,amsfonts}
\usepackage{xcolor}
\usepackage{graphicx} 
\usepackage{todonotes}
\usepackage{acro}
\usepackage{url}
\usepackage{siunitx}
\usepackage{listings}
\usepackage[hidelinks]{hyperref}
\usepackage{subfigure}

\usepackage{xspace}
\let\OldTexttrademark\texttrademark
\renewcommand{\texttrademark}{\OldTexttrademark\xspace}%

\usepackage{booktabs}
\usepackage{tabularx}
\newcolumntype{Y}{>{\centering\arraybackslash}X}

\usepackage{tikz}
\newcommand*\circled[1]{\tikz[baseline=(char.base)]{
            \node[shape=circle,draw,inner sep=0.1ex] (char) {#1};}}

\DeclareSIUnit{\tbps}{Tb/s}
\DeclareSIUnit{\tb}{Tb}
\DeclareSIUnit{\gbps}{Gb/s}
\DeclareSIUnit{\gb}{Gb}

\def\BibTeX{{\rm B\kern-.05em{\sc i\kern-.025em b}\kern-.08em
    T\kern-.1667em\lower.7ex\hbox{E}\kern-.125emX}}

\usetikzlibrary{svg.path}
\definecolor{orcidlogocol}{HTML}{A6CE39}
\tikzset{
    orcidlogo/.pic={
            \fill[orcidlogocol] svg{M256,128c0,70.7-57.3,128-128,128C57.3,256,0,198.7,0,128C0,57.3,57.3,0,128,0C198.7,0,256,57.3,256,128z};
            \fill[white] svg{M86.3,186.2H70.9V79.1h15.4v48.4V186.2z}
            svg{M108.9,79.1h41.6c39.6,0,57,28.3,57,53.6c0,27.5-21.5,53.6-56.8,53.6h-41.8V79.1z M124.3,172.4h24.5c34.9,0,42.9-26.5,42.9-39.7c0-21.5-13.7-39.7-43.7-39.7h-23.7V172.4z}
            svg{M88.7,56.8c0,5.5-4.5,10.1-10.1,10.1c-5.6,0-10.1-4.6-10.1-10.1c0-5.6,4.5-10.1,10.1-10.1C84.2,46.7,88.7,51.3,88.7,56.8z};
        }
}
\newcommand\orcidicon[1]{\href{https://orcid.org/#1}{%
        \mbox{\begin{tikzpicture}[yscale=-1,transform shape, scale=0.03] 
                \pic{orcidlogo};
            \end{tikzpicture}}}}

\usepackage{fancyhdr}
\fancypagestyle{IEEEtitlepagestyle}{
    \fancyhf{} 
     
    \fancyhead[C]{
        \small This work has been accepted at the \emph{4th KuVS Workshop on Network Softwarization (KuVS NetSoft)} under the Creative Commons Attribution 4.0 International License (CC BY 4.0).
        \textit{Digital Object Identifier \href{http://dx.doi.org/10.15496/publikation-105108}{10.15496/publikation-105108}}
        }
}

\input{macros.tex}
\input{acronyms.tex}

\begin{document}

\title{\ac{rbfrt}: Fast Runtime Control for the Intel Tofino}

\author{\IEEEauthorblockN{
        Etienne Zink$^{\orcidicon{0009-0001-0879-535X}}$,
        Moritz Flüchter$^{\orcidicon{0009-0006-6047-5827}}$,
        Steffen Lindner$^{\orcidicon{0000-0002-5274-4621}}$,
        Fabian Ihle$^{\orcidicon{0009-0005-3917-2402}}$,
        Michael Menth$^{\orcidicon{0000-0002-3216-1015}}$}
    \IEEEauthorblockA{University of T\"ubingen, Chair of Communication Networks
        \\Email: \{etienne.zink, moritz.fluechter, steffen.lindner, fabian.ihle, menth\}@uni-tuebingen.de}
    \thanks{The authors acknowledge the funding by the Deutsche Forschungsgemeinschaft (DFG) under grant ME2727/3-1.
        The authors alone are responsible for the content of the paper.}
}

\maketitle

\begin{abstract}
    Data plane programming enables the programmability of network devices with domain-specific programming languages, like P4.
    One commonly used P4-programmable hardware target is the Intel Tofino\texttrademark\ switching ASIC.
    The runtime behavior of an implemented P4 program on Tofino\texttrademark can be configured with shell scripts or a Python library from Barefoot provided with the Tofino\texttrademark.
    Both are limited in their capabilities and usability.
    In this paper, we introduce the \ac{rbfrt}, a Rust-based control plane library.
    The \ac{rbfrt} provides a fast and memory-safe interface to configure the Intel Tofino\texttrademark.
    We showed that the \ac{rbfrt} achieves a higher insertion rate for \acs*{MAT} entries and has a shorter response time compared to the Python library.
\end{abstract}


\input{chapters/01-introduction}
\input{chapters/02-background}
\input{chapters/03-implementation}
\input{chapters/04-evaluation}
\input{chapters/05-conclusion}

\bibliography{bibliography/conferences, bibliography/literature}

\bibliographystyle{ieeetr}

\end{document}

%% file: macros.tex
\newcommand\citeN\cite

\newcommand\fig[1]{Figure~\ref{fig:#1}}

\newcommand{\figeps}[3][]{%
   \begin{figure}[tb!]
      \begin{center}
         \leavevmode
         \parbox[t]{#1}{%
            \resizebox{#1}{!}{\includegraphics{figures/#2}}
         }
         \caption{#3\vspace{-0.2cm}}
         \label{fig:#2}
      \end{center}
   \end{figure}
}

\newcommand{\twosubfigeps}[5]{
   \begin{figure}[t]
      \leavevmode
      \begin{center}
         \subfigure[#2]{
            \label{fig:#1}
            \parbox[t]{0.9\columnwidth}{%
               \resizebox{0.9\columnwidth}{!}{\includegraphics{figures/#1}}
               \vspace{-1cm}
            }
         }
         \subfigure[#4]{
            \label{fig:#3}
            \parbox[t]{0.9\columnwidth}{%
               \resizebox{0.9\columnwidth}{!}{\includegraphics{figures/#3}}
               \vspace{-1cm}
            }
         }
      \end{center}
      \vspace{-0.5cm}
      \caption{#5}
   \end{figure}
}

\newboolean{makevspace}
\newcommand{\cvspace}[1]{%
   \ifthenelse
   {\boolean{makevspace}}
   {\vspace{#1}}
   {}%
}

%% file: acronyms.tex
\acsetup{single}

\DeclareAcronym{SDN}{
    short = SDN,
    long = software-defined networking,
}

\DeclareAcronym{rbfrt}{
    short = RBFRT,
    long = Rust Barefoot Runtime,
}

\DeclareAcronym{MAT}{
    short = MAT,
    long = match-action table,
}

\DeclareAcronym{MAU}{
    short = MAU,
    long = match-action unit,
}

\DeclareAcronym{bfrt}{
    short = BFRT,
    long = Barefoot Runtime,
}

\DeclareAcronym{P4TG}{
    short = P4TG,
    long = P4-based traffic generator,
    single = P4TG
}

\DeclareAcronym{TNA}{
    short = TNA,
    long = Tofino\texttrademark\ Native Architecture,
}

%% file: chapters/01-introduction.tex
\section{Introduction}
P4~\cite{BoDa14} is a domain-specific programming language for data plane programming currently used in academia and industry.
It provides an architecture abstraction that allows developers to define forwarding logic independent of the hardware.
The Intel Tofino\texttrademark switching ASIC is one commonly used P4-programmable hardware.
It can be configured with shell scripts or a provided Python control plane library.
Both approaches have significant shortcomings.
The shell scripts can only write static table entries and cannot react to events.
Python controllers react to events but are based on a slow programming language.
Especially in productive environments, the control plane has to be fast and reliable.

Therefore, we present the \acf{rbfrt}, a Rust-based control plane library for the Intel Tofino\texttrademark.
Rust is an up-to-date, fast, and memory-safe programming language~\cite{Rust}.
The \ac{rbfrt} uses the same protocol for communication with the Tofino\texttrademark as the provided Python library.
\ac{rbfrt} facilitates the development of faster and more reliable control planes for the configuration of Tofino\texttrademark.
Moreover, \ac{rbfrt} features batch configuration which is not supported by the Python library and leads to significantly larger insertion rates for table entries.

%% file: chapters/02-background.tex
\section{Configuring P4 Data Planes}
\label{sec:background}

P4 programs define the data plane, e.g., packet processing logic, for P4-programmable switches, called targets.
Targets can be software-based, like the bmv2~\cite{bmv2-git}, or hardware-based like the Intel Tofino\texttrademark switching ASIC~\cite{Tofino}.
In P4, the packet processing logic is implemented in tables, called \acp{MAT}, and applied through packet header and metadata matching.
Further, the target can define other P4 entities like registers.
While the structure of \acp{MAT} and P4 entities is defined in the data plane, they are configured by the control plane.
A controller is a program that implements this configuration logic.

\begin{figure}[t]
    \centering
    \includegraphics[width=0.71\columnwidth]{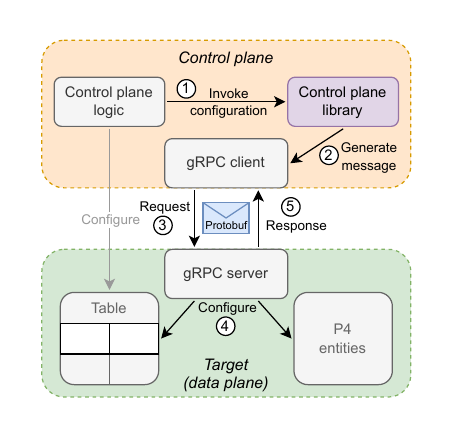}
    \caption{\emph{Configuration of a P4 target by a control plane.
            The control plane logic invokes the control plane library to generate a Protobuf message from the provided data.
            Then, the control plane library sends the Protobuf message with a gRCP client library.
            The target's gRPC server receives the message, and the respective P4 entities, like tables, are configured.}}
    \label{fig:p4sdn}
\end{figure}

\fig{p4sdn} shows how the control plane configures a target.
The \emph{control plane logic} defines how the data plane needs to be configured, e.g., what entries to insert.
The communication between the control plane and target is implemented with Protobuf~\cite{Protobuf} messages and gRPC~\cite{gRPC} requests.
Therefore, the control plane implements a gRPC client and the target a gRPC server.
The control plane logic invokes a \emph{control plane library} \circled{1} to configure the target.
The control plane library generates Protobuf messages \circled{2} and sends them to the target with the gRPC client \circled{3}.
The target's gRPC server receives the request and configures \acp{MAT} and other P4 entities accordingly \circled{4}.
Afterward, the gRPC server responds with an acknowledgment \circled{5} of whether the configuration was successful.
The control plane can configure the target and read data from it.
The gRPC server's response returns the data when reading from the target.

The Protobuf messages are defined in specifications like the \emph{P4Runtime}~\cite{P4Runtime} or \emph{\ac{bfrt}}~\cite{OpenTofino}.
We call these specifications \emph{runtimes} in the following.
The definition of the runtime messages is independent of the target and P4 program.
To facilitate the controller development, \emph{control plane libraries} are provided.
Control plane libraries generate the Protobuf messages for a specific runtime based on provided data.
Because of the target independence, these control plane libraries can be used to configure all targets implementing the respective runtime.

The Intel Tofino\texttrademark is configured through \ac{bfrt} messages.
Barefoot implemented a Python-based control plane library to facilitate the configuration of Tofino\texttrademark.
The library's usability is limited because of its lack of documentation and the effort required to use it.
Further, the library is not maintained anymore.
APS Networks introduced its own Python control plane library for better documentation and ease of use.
Further, this library is no longer developed, little in use, and the GitHub repository~\cite{bfrt-helper-git} is not well documented.
Therefore, we do not consider the library from APS Networks further in this paper.
In addition, both libraries are Python-based, limiting the resulting controller's speed.
Another option to configure the Tofino\texttrademark is to execute commands directly in its shell or to write and execute simple scripts.
Both require much manual effort, are limited in extensibility, and do not react to events.
Below the line, the current tools to configure the Intel Tofino\texttrademark are all limited in usability and resulting controller's speed.

%% file: chapters/03-implementation.tex
\section{The Rust Barefoot Runtime}

We present the \acf{rbfrt}, which is a Rust-based control plane library implementing the \ac{bfrt} and is available on GitHub~\cite{rbfrt-git}.
It is developed to overcome the limitations of the currently available tools to configure the Intel Tofino\texttrademark.
\ac{rbfrt} provides new functions that simplify the configuration process compared to the existing Python library.
The programming language Rust is used because of its fast execution, memory efficiency and safety, and static type system.
As a result, \ac{rbfrt} facilitates the development of faster and more reliable controllers.
Controller implemented with the \ac{rbfrt} are already used by \ac{P4TG} in \cite{LiHae23, IhZi25}, an MNA prototype in \cite{IhMe24}, and an extension to BIER-TE in \cite{FlLi24}.

With \ac{rbfrt}, one or multiple configuration entries are passed to a function call and then sent within a single gRPC request.
We call this \emph{batch configuration}.
In contrast, the existing Python libraries from Barefoot and APS support only a single configuration entry per gRPC request.
Batch configuration reduces overhead in the presence of many timely related configuration entries.
The objective is a higher configuration rate, leading to a faster control of the target.

%% file: chapters/04-evaluation.tex
\section{Evaluation}

In this section, two control plane libraries are evaluated: the Barefoot Python library provided with the Intel Tofino\texttrademark and the \acl{rbfrt}.
First, we describe the testbed and the experiment.
Afterward, we analyze the results and summarize improvements.


\subsection{Testbed}

Two control paradigms are distinguished: local and remote control.
\fig{pdfs/setup} illustrates the testbed for both control paradigms.
The local controller runs directly on the CPU of the Tofino\texttrademark.
The remote controller runs on a server with Ubuntu 22.04.1, an Intel(R) Xeon(R) Gold 6134 CPU @ 3.20 GHz, four cores, 16 GB of RAM, and a Mellanox(R) ConnectX-5 NIC.
It is connected to the Tofino\texttrademark with a single \qty{100}{\gbps} link.
The local controller has limited performance, while the remote controller has a larger delay for the gRPC requests.

\figeps[0.93\columnwidth]{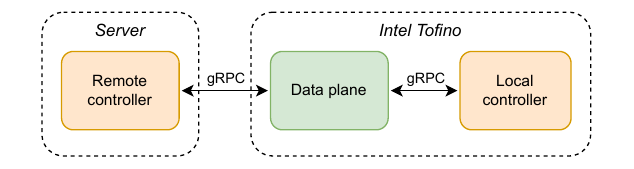}{\emph{Experiment setup with a remote and a local controller. Both communicate via gRPC with the data plane API.}}


\subsection{Insertion of MAT Entries}
\label{sec:insertionMATEntries}

This experiment evaluates the insertion rate and response time for inserting \ac{MAT} entries with different batch sizes.
The insertion rate indicates the number of entries a single controller can insert into a \ac{MAT} over time.
The response time includes the duration for a controller to create and transmit a request until it receives a successful response from the target.
The data plane on the Intel Tofino\texttrademark runs a simple, self-programmed firewall in the experiment.
It consists of one \ac{MAT} that performs an exact match on the source and destination IP address of an incoming packet.
In each evaluation run, the controller inserts $3 \cdot 10^4$ entries into the \ac{MAT} with different batch sizes, i.e., entries per request.
The batch sizes are $\{a \cdot 10^i \mid a \in \{1,3\} \land i \in \{0,1,2,3,4\}\}$.
Single-entry configuration corresponds to a batch size of one, and a batch size of $3 \cdot 10^4$ configures all entries in a single request.
Smaller batch sizes are relevant for productive environments while larger batch sizes are relevant to review the scalability of the control plane library.
The controller creates and sends a batch of configuration entries to the target and waits until the configuration is acknowledged.
Then the next batch is created and sent.
The cumulative time to insert all entries, i.e.,  creation of the first request until the response of the last request, is measured.
The insertion rate and response time are calculated based on this cumulative time.
For any batch size, 100 runs were conducted.
Confidence intervals with an overall significance level of 1\% were calculated.
As the half-widths of all confidence intervals are smaller than 1\% of the measured average, we omit them in the following figures.

\subsection{Results of MAT Insertions}
\label{sec:results}

\fig{pdfs/insertionRate} shows the average insertion rate with different batch sizes.
The local and remote Python controller with single-entry requests achieve a rate of \qty{746}{entries/s} and \qty{553}{entries/s}, respectively.
The \ac{rbfrt} controller with single-entry insertion achieves \qty{1134}{entries/s} and \qty{671}{entries/s}, which is an increase of 48\% and 21\% compared to Python.
Further, the insertion rate of the \ac{rbfrt} increases significantly with the batch size.
This increase results from a lower overhead per entry in larger batches.
With a batch size of $3 \cdot 10^4$ entries, the insertion rate reaches \qty{52832}{entries/s} for the local and \qty{49136}{entries/s} for the remote controller.
Thus, when configuring the Intel Tofino\texttrademark with a single controller, the insertion rate is limited by controller performance and network delay, not by the Tofino\texttrademark itself.
Therefore, larger batch sizes lead to a more efficient insertion of \ac{MAT} entries.

\twosubfigeps{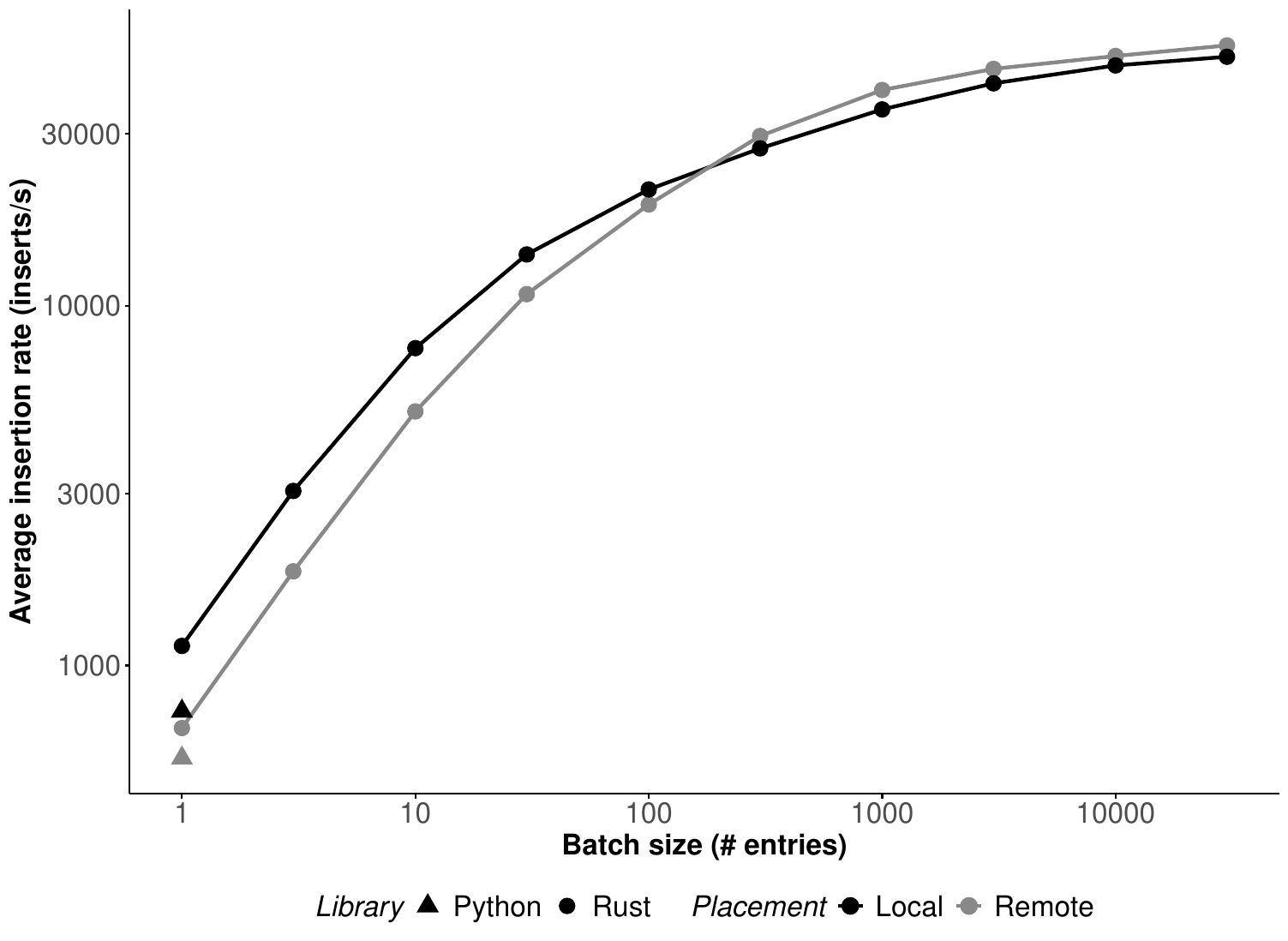}{\emph{Average insertion rate.}}{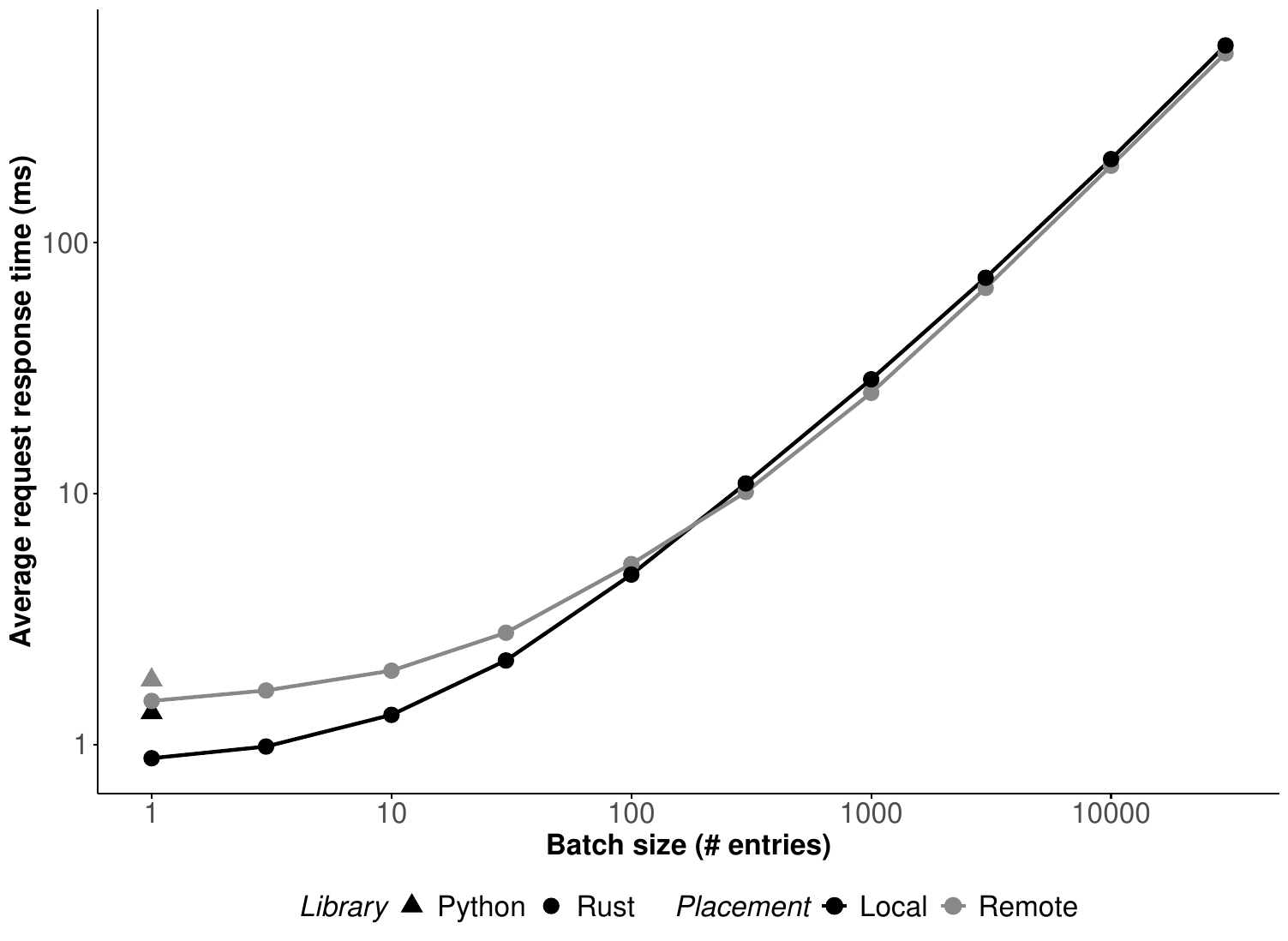}{\emph{Average request response time.}}
{\emph{Evaluation of inserting $3 \cdot 10^4$ entries into the \ac{MAT} of the firewall data plane with different batch sizes, control plane paradigms, and control plane libraries.}}

We further observe in \fig{pdfs/insertionRate} that the insertion rate with small batches is larger for a local than for a remote controller while it is vice-versa for large batches.
This can be explained as follows.
The remote controller runs on more powerful hardware, therefore its insertion rate is better for large batches.
In contrast, small batches induce more network delay per entry, so a better insertion rate is obtained with the lower network delay of the local controller.

\fig{pdfs/requestResponse} shows the average response time for requests with different batch sizes.
Like the insertion rate also the response time increases with larger batch size.
The average response time for the local and remote Python controller is \qty{1.34}{ms} and \qty{1.8}{ms}, respectively.
The corresponding values for the Rust controller (for a single entry) are \qty{0.88}{ms} and \qty{1.49}{ms}, which is a reduction of 34\% and 17\%  compared to Python.
Therefore, an appropriate batch size is recommended so that the insertion rate can be high while the response time for a single request is still low enough.
Given an acceptable response time, a maximum batch size can be derived from \fig{pdfs/requestResponse}.
\fig{pdfs/insertionRate} can be used to deduce the achievable insertion rate for that batch size.

%% file: chapters/05-conclusion.tex
\section{Conclusion}

In this paper, we introduced \ac{rbfrt}, a Rust-based control plane library for the Intel Tofino\texttrademark.
Due to the programming language Rust, the library is fast and memory-safe.
Our experiments showed that \ac{rbfrt} leads to shorter response times for \ac{MAT} entry insertion than the Barefoot Python library, which is shipped with Tofino\texttrademark.
Moreover, \ac{rbfrt} supports batch configurations, which greatly increase achievable insertion rates compared to single-entry insertions offered by the Python library.
While \ac{rbfrt} also supports port configuration and data reading, performance regarding these operations still needs to be evaluated.
Further, for better comparison, batch configurations may be implemented and then evaluated for Barefoot's Python library.